\documentclass[sigconf]{acmart}
\AtBeginDocument{%
  }

\copyrightyear{2026}
\acmYear{2026}
\setcopyright{cc}
\setcctype{by}
\acmConference[AVI '26]{Proceedings of the 2026 International Conference on Advanced Visual Interfaces}{June 08--12, 2026}{Venice, Italy}
\acmBooktitle{Proceedings of the 2026 International Conference on Advanced Visual Interfaces (AVI '26), June 08--12, 2026, Venice, Italy}
\acmDOI{10.1145/3811427.3811444}
\acmISBN{979-8-4007-2342-1/2026/06}
\usepackage{multicol}
\usepackage{soul}    
\usepackage[most]{tcolorbox}  
\tcbuselibrary{skins}
\tcbset{enhanced,colframe=red!50,colback=yellow!25,
nobeforeafter,shrink tight,left skip=1ex, right skip=1ex, extrude by=1.2mm}
\definecolor{shadecolor}{rgb}{.4, .6, .3}
\usepackage{fancybox}
\usepackage{framed}
\usepackage{textcomp}
\usepackage{fontspec}
\usepackage[normalem]{ulem} 
\usetikzlibrary{patterns}
\usepackage{subcaption}
\definecolor{blue}{rgb}{0.36, 0.78, 0.95}
\definecolor{green}{rgb}{0.29, 0.99, 0.40}
\definecolor{gray}{rgb}{0.80, 0.80, 0.80}
\definecolor{fontgreen}{rgb}{0.31, 0.65, 0.18}

\usepackage{graphicx}
\usepackage{float}
\usepackage{enumitem}
\usepackage{xspace}
\usepackage{makecell}
\usepackage{ragged2e}
\usepackage{multirow}
\newcolumntype{L}{>{\RaggedRight\arraybackslash}X}
\newlist{tabitemize}{itemize}{1}
\setlist[tabitemize]{
    label=\textbullet,       
    nosep,                  
    leftmargin=1.2em,       
    before=\vspace{2pt},    
    after=\vspace{-2pt}     
}

\lstdefinestyle{JSON}{
  basicstyle=\ttfamily\small,
  commentstyle=\color{gray},
  keywordstyle=\color{blue},
  stringstyle=\color{red},
  breaklines=true,
  breakatwhitespace=true,
  tabsize=2,
  showstringspaces=false
}

\usepackage[utf8]{inputenc}
\usepackage{geometry}
\usepackage{array}
\usepackage{tabularx}
\usepackage[export]{adjustbox} 
\usepackage[table]{xcolor}
\renewcommand{\tabularxcolumn}[1]{m{#1}}
\usepackage{microtype}
\usepackage{longtable}

\newcommand{\systemname}{%
  {\fontfamily{lmr}\selectfont \textsc{S-prism}}\xspace%
}

\begin{document}
\title{Semantic Prompting: Agentic Incremental Narrative Refinement through Spatial Semantic Interaction
}

\author{Xuxin Tang}
\orcid{0009-0004-7997-6280}
\affiliation{%
  \institution{Virginia Tech}
  \city{Blacksburg, VA}
  \country{United States}}
\email{xuxintang@vt.edu}

\author{Ibrahim Tahmid}
\orcid{0000-0001-8727-4896}
\affiliation{%
  \institution{Virginia Tech}
  \city{Blacksburg, VA}
  \country{United States}}
\email{iatahmid@vt.edu}

\author{Eric Krokos}
\orcid{0000-0003-1350-5297}
\affiliation{%
  \institution{Department of Defense}
  \country{United States}}
  \email{ericpkrokos@gmail.com}

\author{Kirsten Whitley}
\orcid{0000-0003-1356-326X}
\affiliation{%
  \institution{Department of Defense}
  \country{United States}}
  \email{visual.tycho@gmail.com}

\author{Xuan Wang}
\orcid{0000-0002-1381-8958}
\affiliation{%
  \institution{Virginia Tech}
  \city{Blacksburg, VA}
  \country{United States}}
\email{xuanw@vt.edu}

\author{Chris North}
\orcid{0000-0002-8786-7103}
\affiliation{%
  \institution{Virginia Tech}
  \city{Blacksburg, VA}
  \country{United States}}
\email{north@vt.edu}

\begin{abstract}

Interactive spatial layouts empower users to synthesize information and organize findings for sensemaking. While Large Language Models (LLMs) can automate narrative generation from spatial layouts, current collage-based and re-generation methods struggle to support the incremental spatial refinements inherent to the sensemaking process. We identify three critical gaps in existing spatial-textual generation: interaction-revision misalignment, human-LLM intent misalignment, and lack of granular customization. To address these, we introduce \textit{Semantic Prompting}, a framework for spatial refinement that perceives semantic interactions, reasons about refinement intent, and performs targeted positional revisions. We implemented \systemname to realize this framework. The empirical evaluation demonstrat\-ed that \systemname effectively enhanced the precision of interacti\-on-revision refinement. A user study ($N=14$) highlighted how participants leveraged \systemname for incremental formalization through interactive steering. Results showed that users valued its efficient, adaptable, and trustworthy support, which effectively strengthens human-LLM intent alignment. 

\end{abstract}



\maketitle


\section{Introduction}
Writing narratives for sensemaking is rarely a linear process. Instead, it involves iterative loops between the final narrative stage and prior stages such as information foraging, extraction, structuring, and synthesis \cite{pirolli2005sensemaking}. Imagine writing a literature review: when a new, important paper is identified after a draft is finished, the author must incorporate it into existing categories and map its connections to other works. This requires reassessing prior insights and conclusions, then refining the narrative to integrate the new evidence with coherent logic and proper citations. This complex, iterative cycle facilitates \textbf{incremental formalization}, allowing the user to start informally and add or adjust structure as their understanding evolves \cite{shipman1999formality}.
 
To facilitate sensemaking, interactive spatial layouts allow users to leverage spatial intelligence \cite{hegarty2010components} to externalize cognition by organizing and contextualizing fragmented information within a literal ``\textbf{Space to Think}'' \cite{andrews2010space}. While Large Language Models (LLMs) offer an unprecedented opportunity to automate narrative generation from these layouts, current research relies on two primary refinement methods: prompt adjustment \cite{obeid2020chart, fu2025dataweaver} and spatial updates \cite{tang2024steering, tang2025respire, yen2024memolet}. 
However, we identified three critical usability gaps in current methods: \textbf{inter\-action-revision misalignment}, caused by extensive rewording during even minor refinements; \textbf{human-LLM intent misalignment}, stemming from the opaque ``black box'' nature of regeneration; and \textbf{inflexible granular customization}, as results are often formulaic and lack controllable detail.

We propose the \textbf{Semantic Prompting} framework, which leverages a multi-agent pipeline to translate spatial interactions into narrative refinements. Semantic Prompting utilizes an inferring agent to reason about human intent from spatial semantic interactions \cite{endert2012semantic} and steer a downstream refining agent. We implemented the \textbf{S-PRISM} system to realize this framework, which displays intermediate reasoning results and highlights revisions to improve human-LLM intent alignment. Empirical evaluation validates this approach, demonstrating that \systemname substantially enhances refinement precision for targeted positional updates. 
We conducted a user study ($N=14$) to evaluate how \systemname addresses the three identified gaps. The results demonstrate that our spatial interactions are intuitive for steering LLMs and that inference results effectively bridge the gap between human and model intent, enabling incremental refinement of customized narratives. Our analysis of refinement successes and failures provides key insights into the future of semantic prompting.
In summary, this paper makes the following contributions: 

\begin{itemize} 

\item A novel interaction paradigm, \textit{Semantic Prompting}, that bridges spatial semantic interactions with LLM reasoning, enabling targeted narrative revisions through intuitive spatial metaphors.


\item A space-report refinement system, \systemname, powered by a hierarchical multi-agent pipeline. This system instantiates \textit{semantic prompting} to support iterative, human-in-the-loop refinement workflows.

\item An empirical evaluation demonstrating that \systemname significantly outperforms regeneration baselines in refinement accuracy and semantic fidelity.

\item A user study with 14 participants validating the system's efficacy in resolving  three open usability gaps.
\end{itemize}

\section{Related Work}
\newcolumntype{Y}{>{\centering\arraybackslash}X} 
\renewcommand{\tabularxcolumn}[1]{m{#1}} 

\begin{table*}[htbp]
    \centering
    \caption{Comparisons between different LLM writing assistants for refinement.}
    \scriptsize
    \vspace{-10pt}
    \label{tab:com}
    \begin{tabularx}{\textwidth}{
        >{\centering\arraybackslash\bfseries}m{2.2cm} 
        Y Y Y Y  
    }
        \toprule 
        \rowcolor{gray!10} 
        Refinement Mechanisms & 
        \textbf{Regeneration from Scratch}  \includegraphics[height=1.5cm, valign=m, margin=0pt 6pt 0pt 6pt]{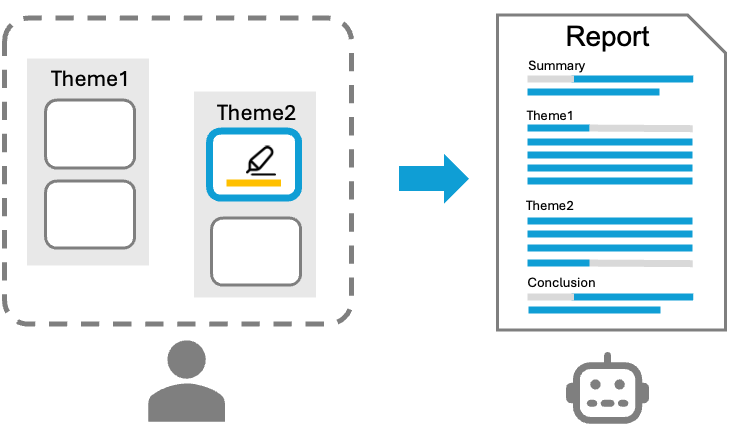} & 
        \textbf{Collage-Based Refinement}  \includegraphics[height=1.5cm, valign=m, margin=0pt 6pt 0pt 6pt]{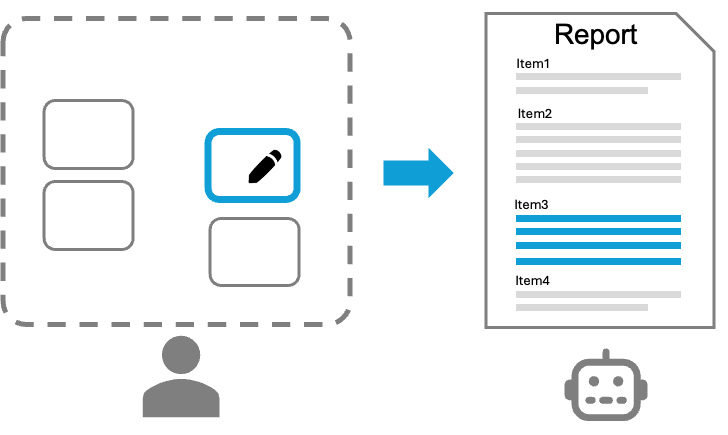} & 
        \textbf{Prompt-Based Refinement}  \includegraphics[height=1.5cm, valign=m, margin=0pt 6pt 0pt 6pt]{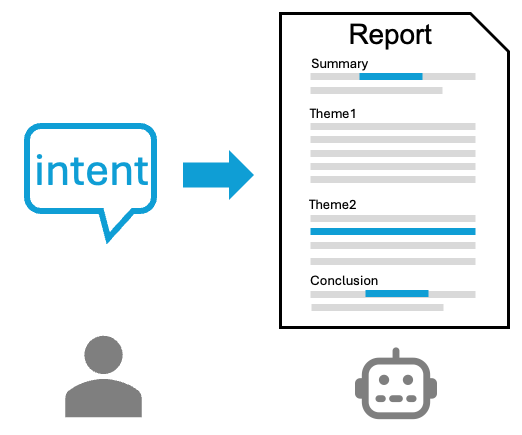} & 
        \textbf{Semantic Prompting} 
        \includegraphics[width=3.2cm, valign=m, margin=0pt 6pt 0pt 6pt]{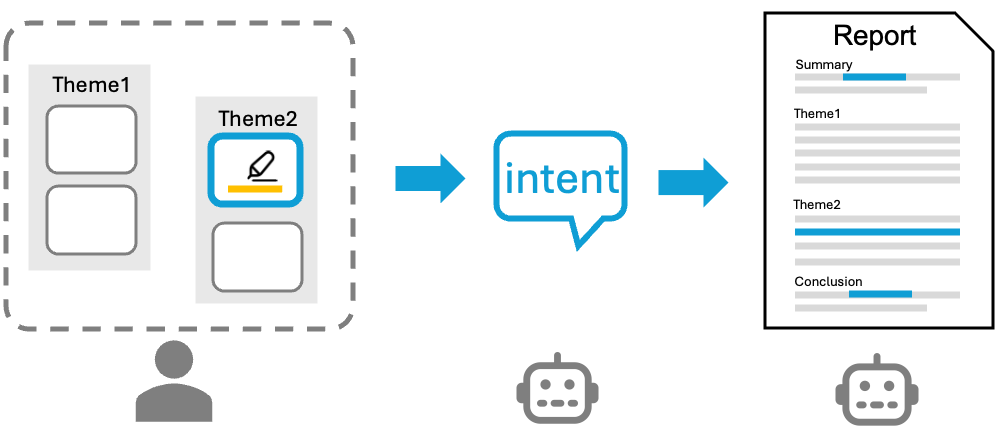} \\ 
        \midrule


        Interactions & Spatial Interactions & Collage-based Interactions & Natural Language Prompts & Spatial Interactions \\ 
        \midrule

        Generation Mechanisms & 
        Regenerate the report for the current space information. & 
        Regenerate only the edited collages. & 
        Manual prompting for expressing explicit revision intents. & 
        LLM-supported reasoning and revision based on semantic interaction.
        \\ 
        \midrule

        Benefits 
        & 
        Update full content with spatial refinements. & 
        Direct object-to-text mapping for revisions.
        & 
        Enable targeted, positional revisions.
        & 
        Update targeted, positional revisions via spatial refinements.
        \\ 
        \midrule

        Drawbacks 
        & 
        Heavily reworded with unintended revisions.
        & 
        Unsuitable for scenarios without direct mapping.
        & 
        Requires proficiency in prompt and context engineering.
        & 
        Nuanced human-LLM intent alignment (Sec~\ref{sec:hl_align})
        \\ 
        \midrule
        
        References  & \cite{tang2025respire, yen2024memolet,tang2024steering}
        & \cite{zhang2023visar, buschek2024collage, cao2025compositional, biermann2022tool, kim2023cells, dang2022beyond, chung2022talebrush}
        & \cite{chatgpt4, anthropic2023claude, du2022read} & \systemname (Our method) \\ 
        \bottomrule 
    \end{tabularx}
\end{table*}

\subsection{Sensemaking, Spatial Layout, and Semantic Interaction}
\textbf{Sensemaking} is a cognitively demanding process where analysts organize fragmented information to uncover insights \cite{pirolli2005sensemaking}. Research has shown that spatial layouts facilitate this process by leveraging visual working memory \cite{andrews2010space, wickens2005design}, a principle now widely adopted by modern collaborative platforms like Miro and Microsoft Teams \cite{miro2025, microsoftteams2025}. While these environments allow users to externalize their "\textbf{Space to Think}" \cite{andrews2010space, shipman1999formality}, the manual effort required to structure information remains significant. To address this, \textbf{Semantic Interaction (SI)} \cite{endert2012semantic} emerged as a paradigm to steer computational models through intuitive spatial actions rather than complex parameter tuning.

However, traditional SI methods are limited by their interpretative depth, relying on simple heuristics \cite{bradel2014multi}, term-weighting \cite{dowling2019interactive}, or expensive fine-tuning \cite{bian2021deepsi} that often fail to capture high-level intent. While modern LLMs offer transformative reasoning capabilities via techniques like Chain-of-Thought \cite{wei2022chain}, they are rarely synchronized with the incremental formalization of spatial layouts. Consequently, we propose \textbf{Semantic Prompting}. Grounded in SI principles, our approach extracts semantic interactions from evolving layouts and translates them—via an LLM inference agent—into inferred prompts, allowing LLMs to dynamically guide downstream analytical tasks.

\subsection{Interactive LLM Writing Assistants}
LLMs are found beneficial for writing assistance \cite{li2024value}. A well-crafted zero-shot prompt can be effective enough for LLM generation \cite{reynolds2021prompt}. 
Despite the prevalence of chatbot interfaces (e.g., ChatGPT and Claude \cite{chatgpt4, anthropic2023claude}), there remains a significant usability gap due to their requirement for prompting expertise \cite{zamfirescu2023johnny}, difficulties with iterative refinements \cite{vasudevan2025llm}, and lack of support for nuanced human control over generation direction \cite{siddiqui2025script}.
Emerging research on interactive LLM writing assistants \cite{radwan2024sard, biermann2022tool, buschek2021impact, chung2022talebrush, dang2023choice, gero2022sparks, gero2023social, goodman2022lampost, kim2023towards, siddiqui2025script, lee2022coauthor, long2023tweetorial, singh2023hide} have bridged this gap.


However, current LLM-based spatial-to-narrative methods struggle to reflect the fluid evolution of a spatial layout, as summarized in Table \ref{tab:com}. Traditional \textbf{regeneration-from-scratch} approaches \cite{tang2024steering, tang2025respire, yen2024memolet} transform spatial context into structured prompts for full report generation, but often suffer from instability and unintended rewording that disrupts incremental formalization \cite{shipman1999formality}. Alternatively, \textbf{collage-based generation} \cite{buschek2024collage, biermann2022tool, zhang2023visar} enables iterative updates via strict object-to-text mapping, yet this paradigm fails when spatial objects lack a direct 1:1 textual correspondence. While \textbf{prompt-based refinement} \cite{du2022read} offers a path for targeted revisions via explicit instructions, efficiently bridging these refinements with intuitive, multi-modal spatial interactions remains an unstudied gap.

We introduce \textbf{Semantic Prompting}, a framework that translates spatial semantic interactions into targeted, positional revisions. Inspired by the ReAct framework \cite{yao2023react} and built upon the ReSPIRE system \cite{tang2024steering}, \systemname utilizes an agentic pipeline to bridge spatial metadata with explicit intent prompts. This architecture shields users from the complexities of manual prompt and context engineering, allowing them to steer narrative refinements directly through intuitive spatial manipulations.


\section{Motivation and Design Goals}
\label{sec:design_goal}






\subsection{Motivation}
\label{sec:motivations}

Spatial layouts serve as a dynamic “\textit{Space to Think}” \cite{andrews2010space}, where the sensemaking process unfolds through incremental formalization \cite{shipman1999formality}. In this process, users do not structure information all at once; instead, they progressively refine their mental models through iterative workspace modifications, such as clustering for semantic proximity and adding visual notes.
However, existing LLM-assisted methods \cite{tang2024steering, tang2025respire, yen2024memolet} struggle to capture these incremental spatial updates, often resulting in a disconnect between the evolving layout and the generated narrative. The absence of a framework to synchronize these layers motivates our subsequent gap analysis ($G$) and design goals ($DGs$).


\subsection{Gap Analysis}
\label{sec:finding}
Specifically, we identify three primary usability gaps arising from the lack of support for incremental processes:

\textbf{G1: Interaction-Revision Misalignment.} 
Current methods like ReSPIRE \cite{tang2025respire} fail to reflect incremental formalism \cite{shipman1999formality} because spatial interactions do not trigger precisely aligned revisions. Instead of targeted updates, the system often regenerates the entire narrative, causing high linguistic variance where significant word-level changes occur despite consistent underlying content. This instability distracts users, forcing them to manually re-verify the integrity of unaffected sections rather than focusing on their intended refinements.

\textbf{G2: Bidirectional Human-LLM Intent Misalignment.}
The “\textit{Two Black Boxes}” problem \cite{wenskovitch2020interactive} fundamentally undermines transparency and explainability in human-LLM collaboration \cite{liao2023ai}, serving as the root cause for the interaction-revision gaps identified in G1. This misalignment is two-fold: an \textbf{Expression Gap}, where users struggle to convey nuanced, multi-level intents—such as distinguishing between a global structural change and a local detail update—through spatial interactions; and a \textbf{Transparency Gap}, where the LLM’s internal reasoning remains opa\-que to the user. For example, a user might highlight a keyword expecting it to be mentioned in the conclusion, but the LLM may respond by merely inserting it into a local paragraph. Without a mechanism to reveal these internal reasoning chains, users cannot verify or correct the system’s inferred intents, hindering the iterative nature of the sensemaking process.

\textbf{G3: Inflexible Granular Customization.}
Although fixed UI designs and backend templates alleviate prompt engineering burdens, they significantly constrain functional flexibility \cite{lehmann2024functional}. Existing spatial-textual systems \cite{zhang2023visar, tang2025respire, chung2022talebrush} often rely on rigid, task-specific mappings that fail to translate nuanced workspace semantics into specific narrative updates. These templatized tra\-nsformations cannot reflect incremental spatial refinements with precision, resulting in a disconnect between the user's externalized cognition and the final text. Addressing this requires a new interaction paradigm that moves beyond one-size-fits-all generation toward individualized, granular revisions that adapt to user-defined contexts across multiple levels of detail.

\subsection{Design Goals}
Grounded in the gaps identified in Section \ref{sec:finding}, we established the following four design goals for the system to support translating incremental spatial refinements into narrative refinements:

\textbf{DG1: Support targeted space-text mapping refinement (G1, G3).} The system should enable partial, incremental, and targeted report revisions that respond directly to semantic interactions \cite{endert2012semantic} within the workspace. This mapping enables personalized customization by translating individual spatial refinements into corresponding narrative revisions.

\textbf{DG2: Surface LLM intent inference (G2).} To ensure transparency and human supervision \cite{liao2023ai, zhang2025neurosync}, the system must present the intermediate inference results of how the LLM reasons about user intent during the refinement process.

\textbf{DG3: Enable user-initiated refinement (G3).} 
To preserve user agency, our system employs a user-initiated mechanism \cite{oppermann2002user}. By delegating the trigger to the user, we ensure that the system only performs report refinements when the user deems their spatial refinements ``ready,'' thereby reducing the cost of correcting premature AI-generated content.



\section{Semantic Prompting and S-PRISM}
\subsection{Semantic Prompting}
\label{sec:s_prompting}
We propose \textbf{Semantic Prompting}, a novel framework that \textit{tra\-nslates users' semantic spatial interactions}—representing their underlying reasoning process—\textit{into executable prompts to steer LLM narrative refinement}.
Grounded on semantic interaction \cite{endert2012semantic}, semantic prompting leverages the cognitive link between a user's spatial interactions and their analytical intent. In this framework, spatial layouts function as an evolving external me\-mory, allowing the AI to interpret incremental, human-led refinements during the sensemaking process.

\begin{figure}[htb]
 \centering
 \setlength{\belowcaptionskip}{-10pt} 
 \includegraphics[width=\columnwidth]{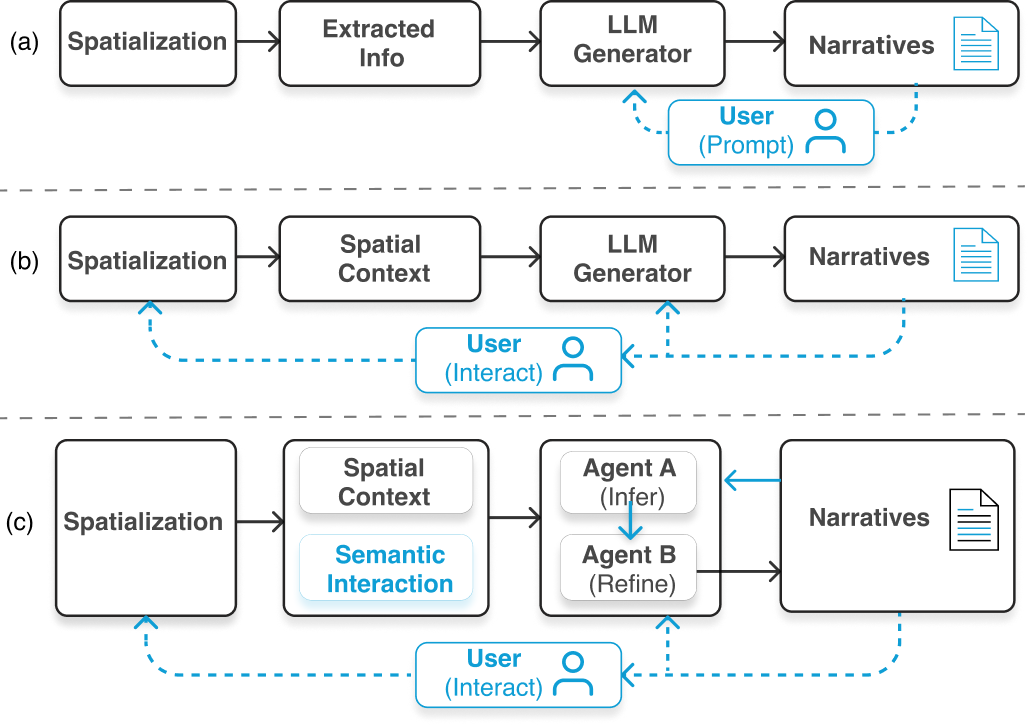}
 \caption{
 Comparison of Space-to-Text Frameworks. (a) Previous methods, such as those in \cite{obeid2020chart, fu2025dataweaver}, center the refinement process on prompt engineering. (b) Space-to-report generation framework \cite{tang2025respire} enables users to  interact directly within a spatial metaphor. (c) Our proposed \textit{Semantic Prompting} framework translates new semantic interactions into revisions, supporting incremental refinement. Blue dashed line: human interactions; blue solid line: our method’s newly-introduced data flow.
 }
 \label{fig:framework}
\end{figure}

Previous Space-to-Text frameworks \cite{obeid2020chart, fu2025dataweaver} (Fig. \ref{fig:framework}a) generally facilitate iterative refinement through manual prompt modification. To improve human-in-the-loop workflows, Tang et al. \cite{tang2025respire} introduced space-to-report generation (Fig. \ref{fig:framework}b), enabling users to drive narrative updates directly via spatialization. However, this method typically regenerates the entire report, which introduces unintended rewording and lacks the granularity required for precise steering.
To bridge this gap, we introduce \textbf{Semantic Prompting}  (Fig. \ref{fig:framework}c), a multi-agent framework designed to enhance refinement precision. Rather than relying on a single LLM generator, our pipeline utilizes specialized agents: an intent inferencer to translate semantic interactions into executable prompts, and a refining agent to perform targeted, positional refinements. This architecture ensures that intuitive spatial manipulations result in precise narrative outcomes, as detailed in Section \ref{sec:system}.

\subsection{S-PRISM System}
\label{sec:system}
\begin{figure}[htb]
 \centering
 \setlength{\belowcaptionskip}{-10pt} 
 \includegraphics[width=\columnwidth]{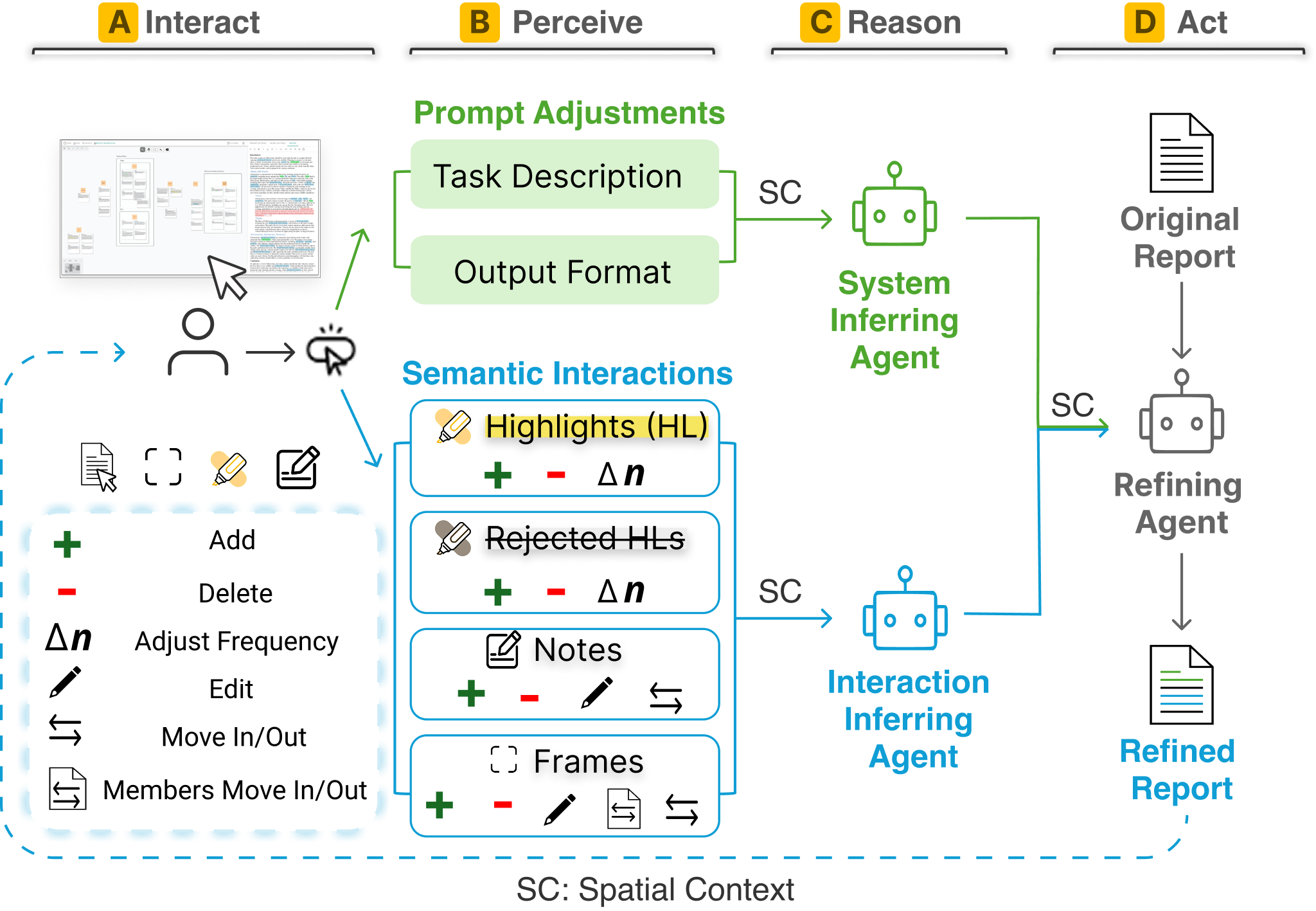}
 \caption{S-PRISM's multi-agent pipeline. Users first interactively refine  the space (A). After triggering the refinement button, S-PRISM perceives the prompt adjustment and semantic interactions (B). In the reasoning step, LLM agents infer the human intents behind the interactions (C). Finally, the refining agent revises the original report based on the reasoned human intents (D). 
 }
 \label{fig:pipeline}
\end{figure}

Built based on ReSPIRE \cite{tang2025respire}, \systemname\footnote{\textbf{S-PRISM} is \textbf{S}emantic  \textbf{P}rompting and \textbf{R}efinement for \textbf{I}nteractive \textbf{S}patial \textbf{M}anipulation} offers two unique advantages: 1) \textit{the Semantic Prompting framework enabling intuitive user-driven LLM refinement}; and 2) \textit{a hierarchical multi-agent pipeline} that manages reasoned semantic prompts and system prompts in the first layer, followed by a second-layer refining agent to act the refinement process. We divide the pipeline into four parts (Fig.~\ref{fig:pipeline}): \textbf{Interact}, \textbf{Perceive}, \textbf{Reason}, and \textbf{Act}. This structure formalizes human-LLM collaboration, enabling the systematic mapping between semantic interactions and the intended LLM-inferred refinement detailed in Table~\ref{tab:interaction-logic}.

\begin{table}[ht]
\centering
\small
\scriptsize
\caption{Mapping between semantic interactions to intended LLM-inferred refinement.}
\vspace{-10pt}
\renewcommand{\arraystretch}{1.2} 
\begin{tabularx}{\columnwidth}{@{} m{1.5cm} X @{}}
\toprule
\multicolumn{1}{c}{\makecell[c]{\textit{\textbf{Form of Semantic}} \\ \textit{\textbf{Interaction}}}} & \multicolumn{1}{c}{\textit{\textbf{Intended LLM-inferred Refinement}}} \\ 
\midrule
\textbf{Text Highlighting} & 
\begin{itemize}[leftmargin=*, nosep, after=\vspace{-\smallskipamount}]
    \item \textbf{Add/Remove:} Incorporate or exclude specific info from the report.
    \item \textbf{Adjust Frequency:} Modify the emphasis or detail level of the text.
    \item \textbf{Rejected Highlights} act as the inverse of standard highlighting.
\end{itemize} \\ \midrule
\textbf{Annotations (Sticky Notes)} & 
\begin{itemize}[leftmargin=*, nosep, after=\vspace{-\smallskipamount}]
    \item \textbf{Add/Remove:} Integrate or delete content associated with the note.
    \item \textbf{Edit/Reposition:} Refine the substance or reassign content to a different narrative block.
\end{itemize} \\ \midrule
\textbf{Document Framing (Clustering)} & 
\begin{itemize}[leftmargin=*, nosep]
    \item \textbf{Add/Remove:} Initialize or delete paragraphs based on the cluster.
    \item \textbf{Rename:} Update headings and edit the summary if necessary.
    \item \textbf{Hierarchy:} Relocate sections and update parent paragraph context.
\end{itemize} \\ \bottomrule
\end{tabularx}
\label{tab:interaction-logic}
\end{table}

\subsubsection{Pipeline}
Fig.~\ref{fig:pipeline} outlines the system pipeline\footnote{All refinement prompts are provided in Supplemental Material A.2.} and the functional role of each component. \systemname enables users to generate initial reports by leveraging spatial context, which is serialized into two primary streams: frame-grouped document contents and interaction metadata, such as highlight frequencies and frame-anchored notes. Once the initial report is established, the interface dynamically transitions to an iterative refinement mode, signaled by the ``\textit{Report Generation}" button updating to ``\textit{Report Refinement}". This state shift allows users to move from bulk generation to precise, interaction-aligned steering of the existing narrative.

\textbf{Interact (Fig.~\ref{fig:pipeline}A).}
\systemname allows users to steer report revisions through a dual-input model consisting of textual inputs and spatial manipulations, which the system translates into refined prompts for the refining LLM. The core of this workflow is grounded in spatial semantic interactions, including: \textit{moving} documents and notes into or out of frames to reorganize structure, \textit{framing} for logical grouping, \textit{highlighting} for modulating the emphasis of specific keywords, and \textit{noting} for attaching contextual human insights. These spatial metaphors—grouping, emphasis, and contextual grounding—are concretely realized through \systemname's frames, highlights, and notes, providing a direct mapping between the spatial layouts and narrative intent. To execute these changes, users click the ``\textit{Report Refinement}'' button, which triggers the refinement (DG3).

\textbf{Perceive (Fig.~\ref{fig:pipeline}B).}
\label{sec: delta_extraction}
After users trigger the refinement, \systemname perceives the two type interactions: prompt adjustments and semantic interactions. Users can adjust the task description and output format of reports for prompt adjustments. Their interactions within the spatial layouts are calculated and perceived as semantic interactions \cite{endert2012semantic}. \systemname includes the perception of 15 types of semantic interactions (Fig.~\ref{fig:pipeline}B), including \textit{addition} and \textit{deletion} (frames, highlights, notes), \textit{edits} (number edits for highlights, textual edits for frames and annotations), and \textit{positional adjustments} (spatial reorganization of the frames, documents, and notes).


\textbf{Reason (Fig.~\ref{fig:pipeline}C).}
The inferring agents in \systemname act as the essential cognitive bridge between raw spatial interactions and structured positional refinements, employing a specialized multi-agent hierarchy to ensure narrative alignment. While the system inferring agent reasons about task-level shifts or formatting adjustments when system prompts change, the interaction inferring agent decodes latent human intent by analyzing perceived semantic interactions—such as clusters, highlights, and notes—within the spatial context. These agents are specifically prompted to address two core questions: (1) \textit{Why did the user perform these interactions?} and (2) \textit{How should the report be revised accordingly?} To fulfill the goal of reasoning transparency (DG2), \systemname surfaces these agentic insights through interface bubbles, allowing users to verify the system’s internal logic before refinements are finalized.

\textbf{Act (Fig.~\ref{fig:pipeline}D).}
Following the inference phase, \systemname consolidates the synthesized intents to guide the refining agent in executing narrative updates. Governed by the principle of interaction-revision alignment (DG1), the agent performs modifications based on three strict heuristics: (1) minimizing unnecessary rephrasing to maintain text stability, (2) restricting the scope of edits to the user’s intended sections, and (3) preserving structural consistency with the output format. To facilitate seamless verification, the system’s backend performs a diff-comparison between the original and revised reports, automatically highlighting the delta in the report panel. This localized highlighting allows users to quickly audit the agent's work, ensuring that refinements remain steerable and transparent without requiring a full-document review\footnote{Usage Examples are provided in Supplemental Material A.1.}.

\subsubsection{Interface and Functionalities}
\begin{figure*}[htbp]
    \centering
    \includegraphics[width=0.8\textwidth]{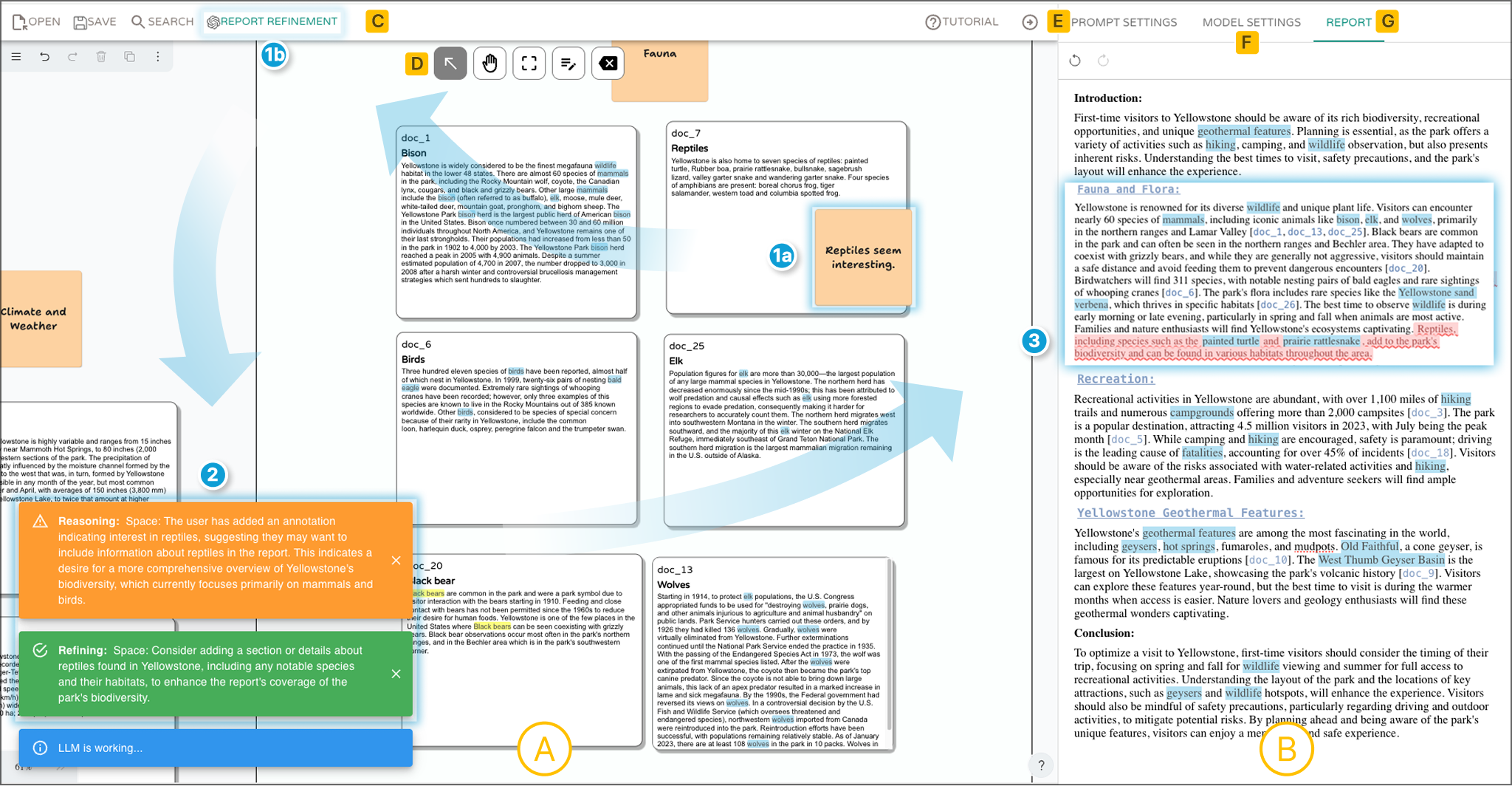}
    \caption{The S-PRISM interface. (A) A direct-manipulation zoomable workspace for spatial document organization and (B) a functional sidebar for system controls. Users manage datasets and snapshots (C) and utilize a toolbar (D) to add frames and notes; document highlighting is performed via double-clicking and the node context menu. The sidebar contains three tabs: (E) prompt settings, (F) model configurations, and (G) the generated report. Within the report, LLM-generated refinements are highlighted in red. We present the front-end steps for the refinement in blue: (1) the user adds a note and triggers the ``\textit{Report Refinement}'' button; (2) reasoning messages by inferring agents; and (3) refined text in red by the refining agent.
    }
    \label{fig:interface}
\end{figure*}
The \systemname interface (Fig.~\ref{fig:interface}) is divided into two main components: the workspace (Fig.~\ref{fig:interface}A) and the sidebar (Fig.~\ref{fig:interface}B). All functionalities are presented in the following sections.

\textbf{Workspace (Fig.~\ref{fig:interface}A).}
The direct-manipulation workspace supports spatial organization through frames, highlights, notes, and an eraser (Fig.~\ref{fig:interface}(C)). It includes flexible navigation (zoom, pan, bird’s-eye view) and undo/redo functionality. Triggering ``\textit{Report Refinement}" (Fig.~\ref{fig:interface}(1)) displays the agent's reasoned intents in orange and narrative revisions in green (Fig.~\ref{fig:interface}(2)) for immediate verification.

\textbf{Sidebar (Fig.~\ref{fig:interface}B).}
The right sidebar provides essential configuration and output controls through three functional tabs: (1) \textbf{Prompt Settings}, which allows users to define task descriptions and customize the report structure (e.g., reordering or renaming components) beyond the default BLUF (Bottom Line Up Front) style; 
(2) \textbf{Model Settings}, for selecting LLM engines (e.g., GPT-4o, GPT-4.1) and fine-tuning parameters like temperature and token limits; 
and (3) \textbf{Report}, a dynamic panel that automatically surfaces generated or refined narratives (Fig.~\ref{fig:interface}) while supporting version control (undo/redo), manual editing, and spatial provenance tracing via direct links back to the workspace.

\subsubsection{Implementation}
\label{sec:imple}
The frontend of \systemname was built with React, tldraw\footnote{https://tldraw.dev/}, and Lexical\footnote{https://lexical.dev/}. Specifically, tldraw powers the wor\-kspace, while Lexical facilitates the interactive reporting features. The backend, implemented in Node.js, manages authentication, logging, and database operations. The system features a model-agnostic multi-agent framework via the OpenAI API, with \textit{GPT-4o-mini} serving as the default model due to its balance of speed and structural reliability in generating the complex structured JSON required for seamless spatial-narrative mapping.

\section{Evaluation}
We standardized on \textit{GPT-4o-mini} across all evaluations to ensure experimental consistency. Although \systemname is model-ag\-nostic, we prioritized this model for its low latency and high inference speed, which are essential for maintaining real-time synchronization between spatial clusters and narrative outputs.

\subsection{Empirical Evaluation}
\label{sec:tech_eval}
To evaluate interaction-revision alignment (G1), we compared \systemname with the ``Regeneration-from-Scratch'' baseline. The analysis focuses on main frames, highlights, and notes to account for the baseline’s lack of nested subframes.

To evaluate the alignment efficacy of \systemname, we conducted a quantitative empirical evaluation across two dimensions of alignment: (1) \textbf{Interaction-to-Revision \textit{Targeted Refinement}}, assessing whether anchored revisions precisely target only the corresponding report paragraphs \cite{birnholtz2012tracking}; and (2) \textbf{Revision-to-Inte\-raction \textit{Semantic Fidelity}}, measuring how faithfully the report revisions reflects semantic interactions made in the worksp\-ace \cite{shipman1999formality}. We measured Precision, Recall, and F1-score across a test dataset of 35 experimenter-curated pairs\footnote{The details are provided in Supplemental Material A.3.} derived from \textit{Sign of the Crescent}, 
a standard sensemaking benchmark. From this corpus, two distinct 10-document subsets were selected to simulate complex, high-information-density scenarios; while professional in nature, these materials were used to test general sensemaking logic without requiring specialized intelligence expertise. We compared \systemname against a regeneration-from-scratch baseline (ReSPIRE \cite{tang2025respire}). As detailed in A.3 of the Supplemental Materials, the dataset covered 14 interaction combinations—ranging from word-level highlights to paragraph-level reorgani\-zations—to rigorously test the system's ability to maintain intent across varied granularities.

\subsubsection{Targeted Refinement}
Following a Bottom Line Up Front (BLUF) structure where each body paragraph maintains a strict one-to-one mapping with a main frame, we evaluated targeted refinement through spatial projection. We define $N_{TPP}$ as the number of correctly refined paragraphs, $N_{PP}$ as the total number of system-refined body paragraphs, and $N_{TP}$ as the number of target paragraphs that should be refined. Based on these metrics, we calculate precision, recall, and F1 score to measure the correctness and completeness of the system's localized updates.
\vspace{-0.5em}
{\footnotesize
\begin{equation} 
\begin{aligned}
Precision_{tr} &= N_{TPP}/N_{PP} \\
Recall_{tr} &= N_{TPP}/N_{TP} 
\end{aligned}
\end{equation}
}
\vspace{-1.5em}



\subsubsection{Semantic Fidelity}
We evaluated semantic fidelity thro\-ugh two complementary metrics: Precision and Recall. Precision assesses the correctness of the refinements, calculated as the ratio of correctly modified sentences ($N_{TPS}$) to the total number of system-edited sentences ($N_{S}$). Recall measures the completeness of the alignment, calculated as the ratio of realized interactions ($N_{RSI}$) to the total number of semantic interactions ($N_{SI}$).
To identify correctly refined sentences ($N_{TPS}$), we extract specific intent-bearing elements from user interactions to serve as the ground truth for verification: entities and citations from highlights, conceptual keywords from notes, and group names or locations from clusters. These extracted markers act as the objective criteria for determining whether a modified sentence aligns with the source interaction, thereby providing the basis for calculating refinement correctness.
\vspace{-0.5em}
{\footnotesize  
\begin{equation} 
\begin{aligned}
Precision_{sf} &= N_{TPS}/N_S \\
Recall_{sf} &= N_{RSI}/N_{SI}
\end{aligned}
\end{equation}
}
\vspace{-1.5em}


\subsubsection{Result}


As shown in Table~\ref{tab:tech_eval}, \systemname outperforms the Regeneration baseline in precision and F1 score for \textit{Targeted Refinement}. The baseline’s higher recall is a result of its global rewriting approach; conversely, \systemname maintains higher precision by mostly refining correct paragraphs. Regarding \textit{Semantic Fidelity}, both methods show comparable recall, but \systemname demonstrates significantly higher precision and F1 scores. 

\begin{table}[th]
    \centering
    \caption{Comparison of Targeted Refinement and Semantic Fidelity across methods}
    \vspace{-10pt}
    \small
    \scriptsize
    \renewcommand{\arraystretch}{1.1} 
    \begin{tabular}{l|cc|cc}
        \hline
        & \multicolumn{2}{c|}{Targeted Refinement} & \multicolumn{2}{c}{Semantic Fidelity} \\
        Methods & Regeneration & \systemname & Regeneration & \systemname \\
        \hline
        Precision & 0.752 & \textbf{0.951} & 0.348 & \textbf{0.558} \\
        Recall    & 1.000 & 0.831 & 0.694 & 0.684 \\
        F1 score  & 0.858 & \textbf{0.887} & 0.463 & \textbf{0.614} \\
        \hline
    \end{tabular}
    \label{tab:tech_eval}
\end{table}
\subsection{User Study}

We conducted a user study to evaluate \systemname, investigating how effectively users can steer it for report refinement. Our analysis addresses the following research questions, which are grounded in the gap analysis presented in Section~\ref{sec:design_goal}.





\textbf{RQ1:} \textit{How successful do users leverage \systemname to purposely refine their reports through spatial semantic interaction (G1)}?

\textbf{RQ2:} \textit{Do the LLM-reasoned intents align with those of the users, and what characterizes successful versus failed refinements (G2)?}

\textbf{RQ3:} \textit{Does \systemname support granular customization} (G3)?



\subsubsection{Participants}
Fourteen participants (9 F, 5 M; age $M = 27, SD = 3.75$) were recruited through university mailing lists. All participants were proficient in English and reported high familiarity with LLMs such as ChatGPT and Claude. Regarding textual synthesis and reporting for sensemaking, the group included two beginners, two intermediate users, nine proficient users, and one expert; for spatial workspaces, experience levels ranged from no experience (3) and beginner (4) to intermediate (4), proficient (2), and expert (1). The selection criteria prioritized individuals who frequently perform sensemaking tasks while ensuring a diverse range of spatial workspace experience, thereby aligning with our target users: knowledge workers who require expressive steering mechanisms for complex synthesis, regardless of their prior familiarity with spatial interfaces.

\subsubsection{Procedure}

The study was conducted in person within a quiet, controlled environment with an experimenter present. Participants accessed the systems via Google Chrome on an external 55-inch Sony monitor using a hardware keyboard and mouse. After obtaining informed consent and demographic information, the experimenter introduced the system’s features through a structured walk-through tutorial. 
The evaluation was structured into two phases\footnote{The prompts for two phases are available in Supplemental Materials A.4.} utilizing distinct datasets and task clusters: (1) closed-ended tasks focused on sequential refinement via semantic interaction exclusively, and (2) open-ended tasks involving purpose-shifted refinement through the integration of semantic interaction and prompt adjustment.
The study was conducted in two phases over approximately 90 minutes, with participants compensated \$20 USD. In Phase I, participants performed four sequential guided report-refinement tasks (Fig.~\ref{fig:p1_task} ($T_{1}$–$T_{4}$)) using the \textit{Sign of the Crescent} dataset 
to maintain consistency with the empirical evaluation. Following an interim questionnaire, Phase II transitioned to open-ended synthesis where participants managed three travel-planning scenarios (Fi\-g.~\ref{fig:p04_shift} ($S_{1}$–$S_{3}$)) using a Yellowstone National Park corpus \cite{wiki:yellowstone_climate}. To isolate the evaluation of the refinement process from initial document triage, these documents were spatially pre-clustered with category-defining notes. The procedure concluded with a post-study questionnaire and a semi-structured interview regarding the participants' overall experience.

\subsection{Results}

The results comprise an analysis of both quantitative log data and qualitative interview insights. Our findings address the three research questions as follows:

\begin{figure}[htbp]
    \centering
    \includegraphics[width=\columnwidth]{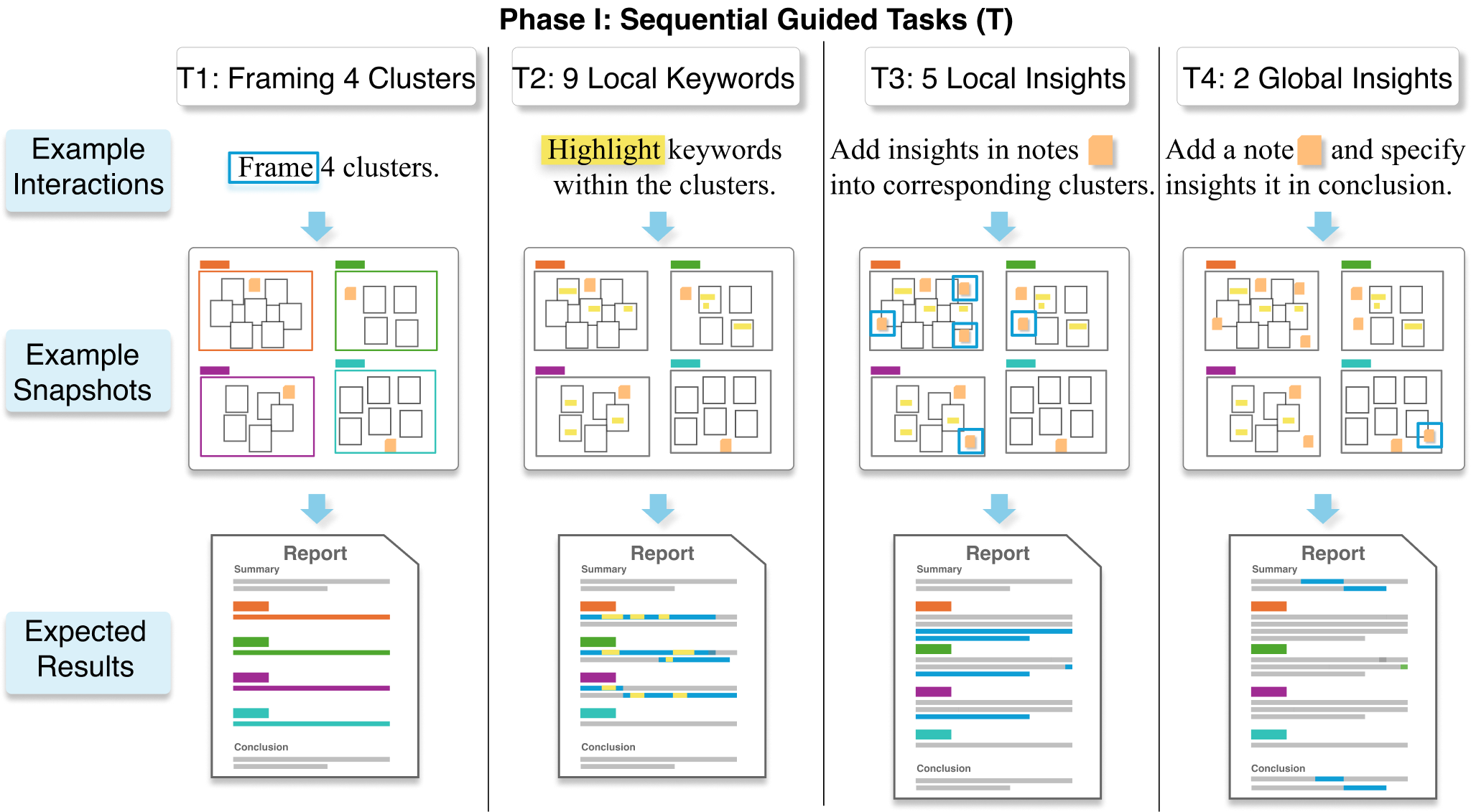}
    \caption{Phase I: An overview of the four tasks (T), illustrated via sample interactions, schematic sketches of snapshots, and expected reports. \systemname ensures the final report features a logical structure, precise insights, and highlighted keywords grounded in objective evidence.
    }
    \label{fig:p1_task}
\end{figure}

\begin{figure}[htbp]
    \centering
    \includegraphics[width=\columnwidth]{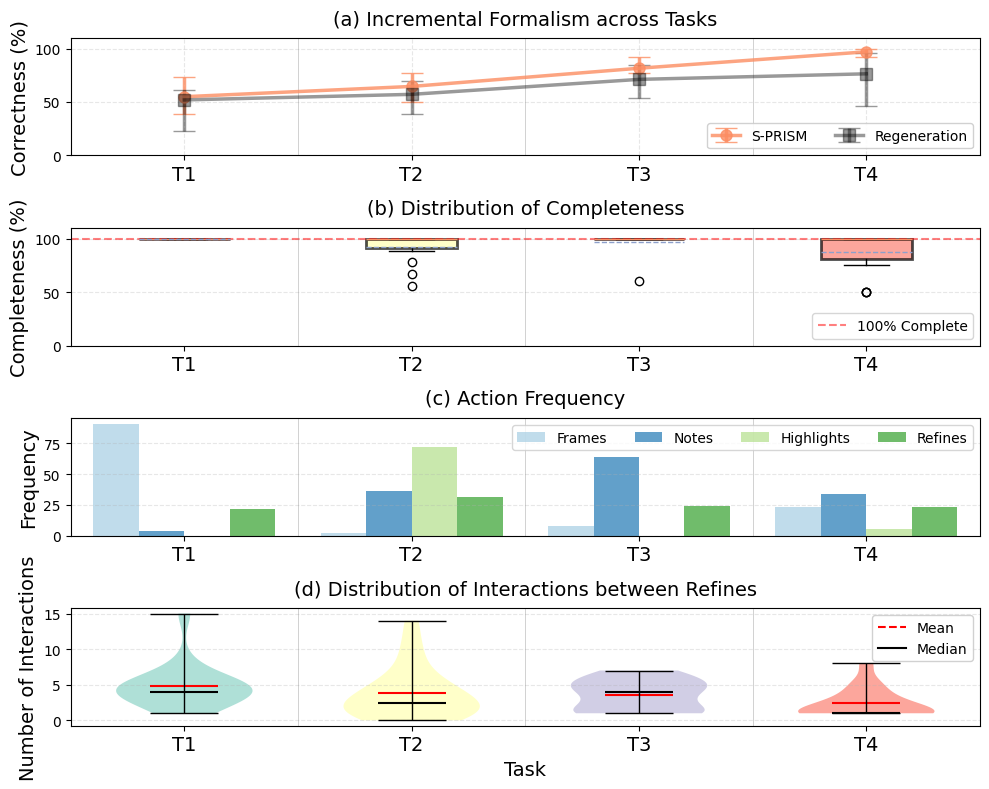}
    \caption{Phase I Behavior Patterns. (a) Average correctness and score distribution (min-max) of final reports by task, illustrating the steady progression of incremental formalism. (b) Task completeness scores, highlighting variations in execution status. (c) Distribution of interaction frequencies by action type, revealing shifts in user strategies and engagement patterns across different tasks. (d) Distribution of spatial interaction frequency per refinement cycle across tasks $T_1$–$T_4$.
    }
    \label{fig:log}
\end{figure}


\subsubsection{Interaction-Revision Alignment (RQ1)}
Following the technical validation of improved precision in interaction-revision alignment (Section~\ref{sec:tech_eval}), we conducted Phase I in the user study to evaluate how participants steer \systemname for incremental refinement. We established a grading rubric and implemented a lexical algorithm to objectively measure report correctness, wh\-ere a score of 100\% denotes the successful completion of all task requirements. To visualize the efficacy of user-driven revisions, we plotted an incremental correctness curve. As a post-hoc baseline, we generated reports from scratch using a consistent model and system prompt across the respective participant spaces for each task. This allows us to evaluate the incremental improvement of \systemname by comparing its output against the Regeneration baseline.

\textbf{Incremental Formalism.}
Results in Fig. \ref{fig:log}a demonstrate that \systemname \textbf{effectively supports incremental formalism}, promoting a steady, cumulative improvement in report correctness throughout the refinement process. In contrast, the Regeneration baseline fails to show consistent growth; this is particularly evident in $T_4$, where average correctness plateaus and the variance (min-max range) expands, indicating high instability in complex scenarios. Qualitative analysis reveals that while \systemname allows participants to incrementally integrate information via spatial layouts, the Regeneration method often produces reports that aggregate or omit user inputs opaquely. Evidence from $T_4$ highlights this performance gap: while P1 captured every required keyword using \systemname, the baseline omitted nearly half (4 out of 9) as new information overshadowed previous content.

\begin{figure}[htbp]
    \centering
    \includegraphics[width=\columnwidth]{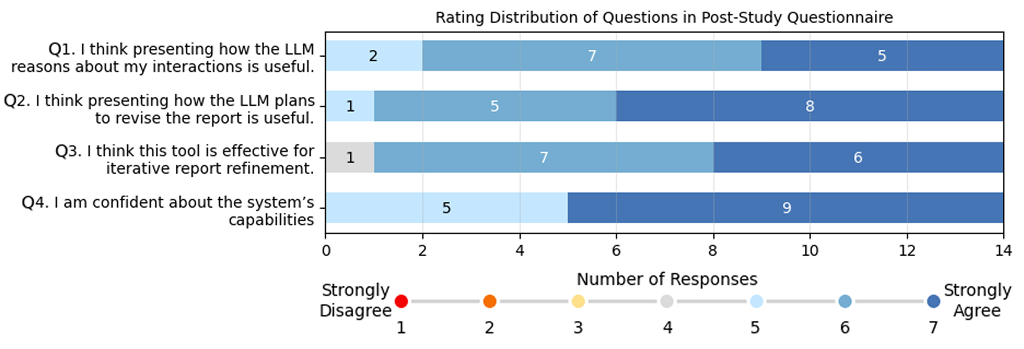}
    \caption{Subjective Ratings. 7-point Likert scores for LLM transparency (reasoning/planning), refinement effectiveness, and system confidence.
    }
    \label{fig:questionnaire}
\end{figure}

\textbf{User Refining Experience.}
Participants reported \textbf{high utility and efficiency} regarding \systemname's interaction-revision alignment. This is supported by the questionnaire results, where thirteen out of fourteen participants found the tool effective for iterative refinement (Fig.~\ref{fig:questionnaire}(Q3)). Qualitatively, users preferred spatial interactions over traditional chat-based prompting; as P14 noted, “\textit{It’s easier to change the reports just by [using] notes rather than retyping everything in chatbots.}” Even though rated Q3 as neural, P7 still specifically highlighted the efficiency gains, observing that the system removed the “\textit{burden of manually consolidating}” disparate notes and frames, which bolstered overall user confidence in the system's capabilities (Fig.~\ref{fig:questionnaire}(Q4)).

\subsubsection{Human-LLM Intent Alignment (RQ2)}
\label{sec:hl_align}
We assessed intent alignment through subjective satisfaction ratings and log analysis of successful versus failed refinement cycles. Participants evaluated whether the reasoning messages (Fig. \ref{fig:interface}(2)) accurately captured their spatial intents. 

\textbf{User Feedback.}
Participants unanimously agreed that presenting LLM reasoning and revision plans was essential for alignment (Fig.~\ref{fig:questionnaire}(Q1, Q2)) and expressed high overall confidence in the system (Fig.~\ref{fig:questionnaire}(Q4)). Qualitatively, these messages bridged the \textbf{Transparency Gap (G2)} by clarifying the logical chain from spatial interactions to narrative updates. While twelve participants valued the explicit explanatory layer for building trust, P4 and P5 found that highlighting differences within the final report provided sufficient verification.


\textbf{Failed Refinements.}
We measured completeness (Fig. 5b) by the inclusion of 4 clusters ($T_1$), 9 keywords ($T_2$), 5 local insights ($T_3$), and 2 global insights ($T_4$). While $T_1$ and $T_3$ performance was high, $T_2$ and $T_4$ revealed misalignments between system logic and user expectations. In $T_2$, lower completeness for P4, P12, and P13 stemmed from a ``\textbf{logic mismatch}'': while \systemname captures the content of highlights, it treats them as global emphasis cues without positional metadata, contradicting users’ intuitive “\textit{spatial projection}” logic that highlights with\-in a frame would map to specific paragraphs. Notably, other participants (P5-9, P14) successfully mitigated this limitation by using Notes as explicit rectifications to anchor keywords to specific sections. In $T_4$, P2 and P6 experienced “\textbf{attention dilution}” when incorporating global insights via major structural reorganization; the LLM prioritized the complex restructuring over localized content updates, leading to omissions. 

\textbf{Successful Refinements.}
Compared to highlights, the use of frames and notes demonstrated higher success rates, with notes proving critical for overall task completion (Fig.~\ref{fig:log}c). Specifically, frames consistently aided users in defining paragraph structures ($T_1$), while notes—leveraged alongside positional met\-adata—enabled the precise insertion of insights into targeted sections ($T_3$). Our study revealed that notes functioned as “\textbf{cont\-extual meta-prompts}” beyond our initial expectations. Participants employed them for two primary purposes: (1) to explicitly \textbf{rectify refinement errors} caused by the lack of positional information in highlights (P5–9, P14), and (2) to serve as \textbf{concise directives for complex tasks}, such as integrating specific insights into a conclusion ($T_4$) or executing other cross-paragraph revisions. Notably, successful recoveries from failed refinements often followed an “\textbf{iterative probe-and-rectify}” strategy to address \textbf{Expression Gap (G2)}: participants first observed the Agent’s inferred intent; if the output was unsatisfactory, they would escalate to notes containing more specific instructions. For instance, when P6 found that highlights alone failed to place keywords in the correct section, she inserted a note within the corresponding frame specifying that the keyword “\textit{is missing in the generated paragraph.}” This \textbf{User Agency} is quantified in Fig. \ref{fig:log}d, which shows participants performed multiple spatial interactions (median 3–5) to “\textit{batch}” their intent before manually triggering a refinement. By decoupling spatial exploration from LLM execution, \systemname ensures users maintain sovereign control over the refinement rhythm, fostering high system confidence (Fig.~\ref{fig:questionnaire}(Q4)).

\textbf{Spatial Interaction vs. Prompting?}
Contrary to expectations, four participants (P6, P10, P11, P14) successfully completed Phase II tasks—transitioning from comprehensive reports to itin\-eraries—without modifying their textual prompts. We found that the pre-defined templates provided a stable ``\textit{Intent Baseline}," allowing users to steer the Agent solely through spatial layouts and notes. As P10 noted: “\textit{The context of my travel plan [in the space] is enough for the LLM.}”
While this ``\textbf{zero-prompt}" transition is contingent upon the clarity of the spatial signal, it validates the Agent's ability to derive complex operational intents by coupling static narrative goals with dynamic spatial semantics. This suggests that a well-structured workspace can effectively reduce the need for explicit textual re-prompting.

\subsubsection{Granular Customization (RQ3)}
\systemname's support for gr\-anular customization is reflected in both structural and detailed adjustments.
\begin{figure}[htbp]
    \centering
    \includegraphics[width=\columnwidth]{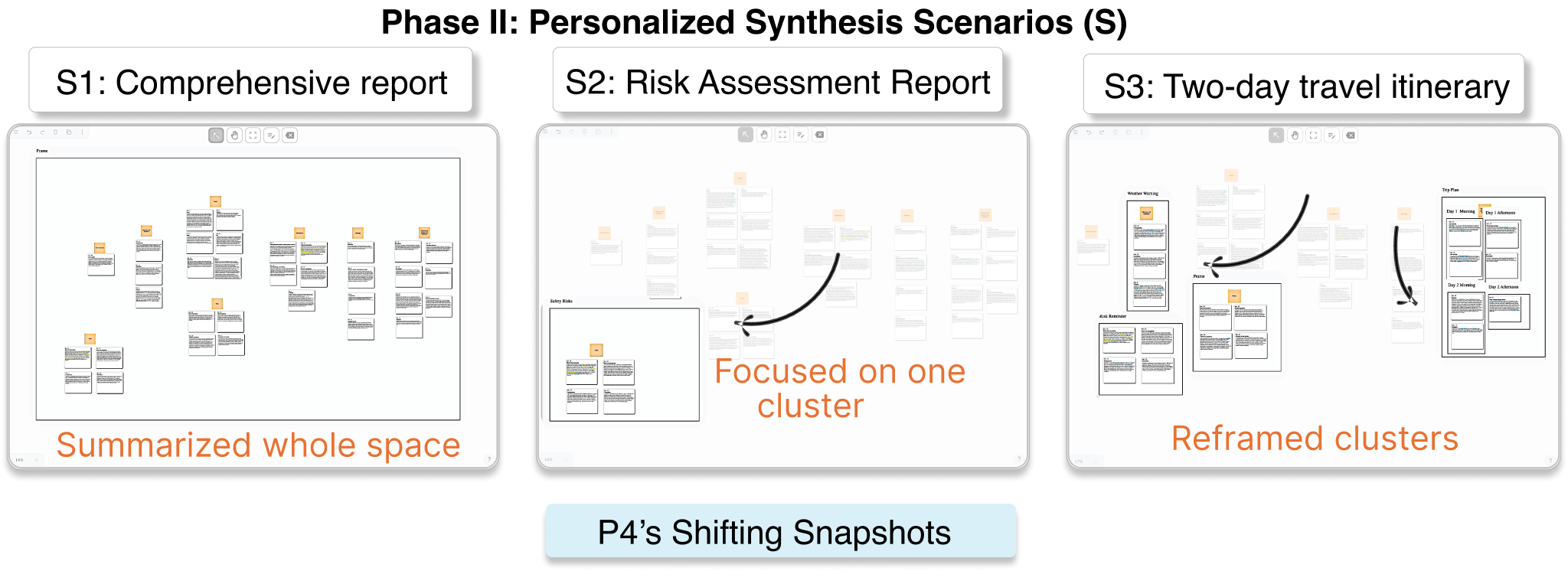}
    \caption{P4 refined the workspace for refining reports across different scenarios in Phase II by creating distinct frames and adjusting prompt settings. Each figure represents a different period: 
    S1, S2, and S3 show the completed workspace snapshots for scenarios 1, 2, and 3, respectively.}
    \label{fig:p04_shift}
\end{figure}

\textbf{Structural Customization.} Participants utilized nested fra\-ming to dynamically reorganize report hierarchies according to shifting analytical goals. 
For example, P4 (Fig. \ref{fig:p04_shift}) moved from a general overview to specific risk clusters, ultimately using sub-frames to distinguish morning and afternoon schedules within a granular two-day itinerary.
This demonstrates how spatial frames in \systemname directly allow users to manage narrative structure at multiple scales.

\textbf{Detail Customization.}
Beyond structure, participants leveraged notes and highlights for precise semantic steering of report content. P1 filtered the Yellowstone corpus through her personal biological lens by emphasizing specific wildlife and flora, while P9 utilized frame-anchored notes to dictate the ``\textit{rhyt\-hm}" of her itinerary—requesting a dense schedule for sightseeing and relaxed intervals for camping. These fine-grained interactions ensure the final output reflects individual analytical priorities and logical preferences.

\section{Discussion}
\subsection{Intent Alignment and Sensemaking Efficacy}
The evaluation results demonstrate that the \textit{Semantic Prompting} framework, as implemented in the \systemname pipeline, successfully enables a reasoning agent to translate spatial semantic interactions into intent-aligned, executable prompts. \systemname substantially enhances refinement precision in empirical tests and, as shown in our user study, steadily elevates report correctness to near-perfect levels, facilitating incremental formalization. Participants provided positive feedback regarding both interaction-revision alignment and human-LLM intent alignme\-nt. These findings indicate that \systemname excels within the iterative sensemaking loop—analyzing spatial modifications within their broader context to execute precise narrative updates.


\subsection{Semantic Prompting for Spatial Writing}
We propose \textit{Semantic Prompting} as a novel paradigm for AI-assisted writing that moves beyond traditional object-text links. Unlike mainstream collage-based methods, this framework enables spatial interactions to transcend local text constraints, facilitating task-specific personalization. The spatial metaphors of grouping, attached insights, and emphasis—realized through \systemname's frames, notes, and highlights—allow for fluid intent expression. Crucially, we found that spatial notes function as meta-prompts to specify or rectify intent. By preserving user agency via the manual ``\textit{Report Refinement}” trigger, \systemname supports a complementary interaction gradient: users can initiate with low-cost highlights and escalate to high-precision notes when complex goals require more than graphical cues.

\subsection{Generalizability}
\systemname supports diverse visual analytics by translating spatial metaphors into domain-agnostic narrative intents. Participants identified potential applications ranging from travel planning to policy investigations. By pairing customizable templates with multi-level control—from word-level highlights to global struct\-ures—\textit{Semantic Prompting} enables precise, context-aware customization across varied tasks.

\section{LIMITATIONS AND FUTURE WORK}
While \systemname aligns spatial intent with report refinement, its limitations offer paths for future research.

\textbf{Technical Scalability and Transparency.} High-density co\-mpound interactions can trigger ``\textit{attention dilution}," where the model prioritizes structural reorganization over specific narrative precision. Despite high user confidence (Fig.~\ref{fig:questionnaire}(Q4)) and appreciation for LLM reasoning (Fig.~\ref{fig:questionnaire}(Q1, Q2)), a residual ``\textit{logic transparency gap}" remains regarding why certain actions fail to trigger revisions. Future work will explore adaptive token management and ``\textit{verify-before-execute}" loops to resolve silent failures and strengthen human-in-the-loop alignment.

\textbf{Workflow and Narrative Management.} \systemname currently lacks spatial reordering, which prevents the layout from functioning as a true visual storyboard. Furthermore, while annotations can serve as meta-prompts to trigger changes across multiple paragraphs, the current spatial-to-report mapping—which relies primarily on spatial clusters—hinders a fluid writing process. Future versions will support more flexible spatial-narrative projections to enable seamless reorganization.

\textbf{Semantic and Mixed-Initiative Alignment.} We aim to shift from Action-level to Goal-level alignment, utilizing multi-stage refinement to distinguish between deep semantic intents like ``\textit{strengthening evidence}" versus ``\textit{refining tone}." Furthermore, we envision mixed-initiative support using ``\textit{graded interventions}," where the system proactively suggests refinements based on spatial conflicts or low-confidence inferences to minimize interaction costs while preserving user agency.

\section{CONCLUSION}
This paper introduces the \textit{Semantic Prompting} framework and its implementation, \systemname, to bridge the gap between spatial semantic interaction and precise narrative refinement. By translating spatial metaphors—such as framing, highlighting, and noting—into intent-aligned prompts via LLM agents, we address critical gaps in interaction-revision alignment, human-LLM intent alignment, and granular customization. Empirical evaluations confirm the higher precision of our interaction-revi\-sion mapping. Furthermore, our user study demonstrates that \systemname effectively addresses the three identified gaps while providing users with efficient, adaptable, and trustworthy support.

\begin{acks}
We would like to thank Rehema Abulikemu for her participation in the initial brainstorming sessions.
    This research was supported by industry, government, and institute members of the NSF SHREC Center, which was founded in the IUCRC program of the National Science Foundation.
\end{acks}
\bibliographystyle{ACM-Reference-Format}
\bibliography{main}

\onecolumn    
\appendix
\section{APPENDIX}
\subsection{S\_PRISM: Examples}
\begin{figure}[H]
    \centering
    \includegraphics[width=0.9\textwidth]{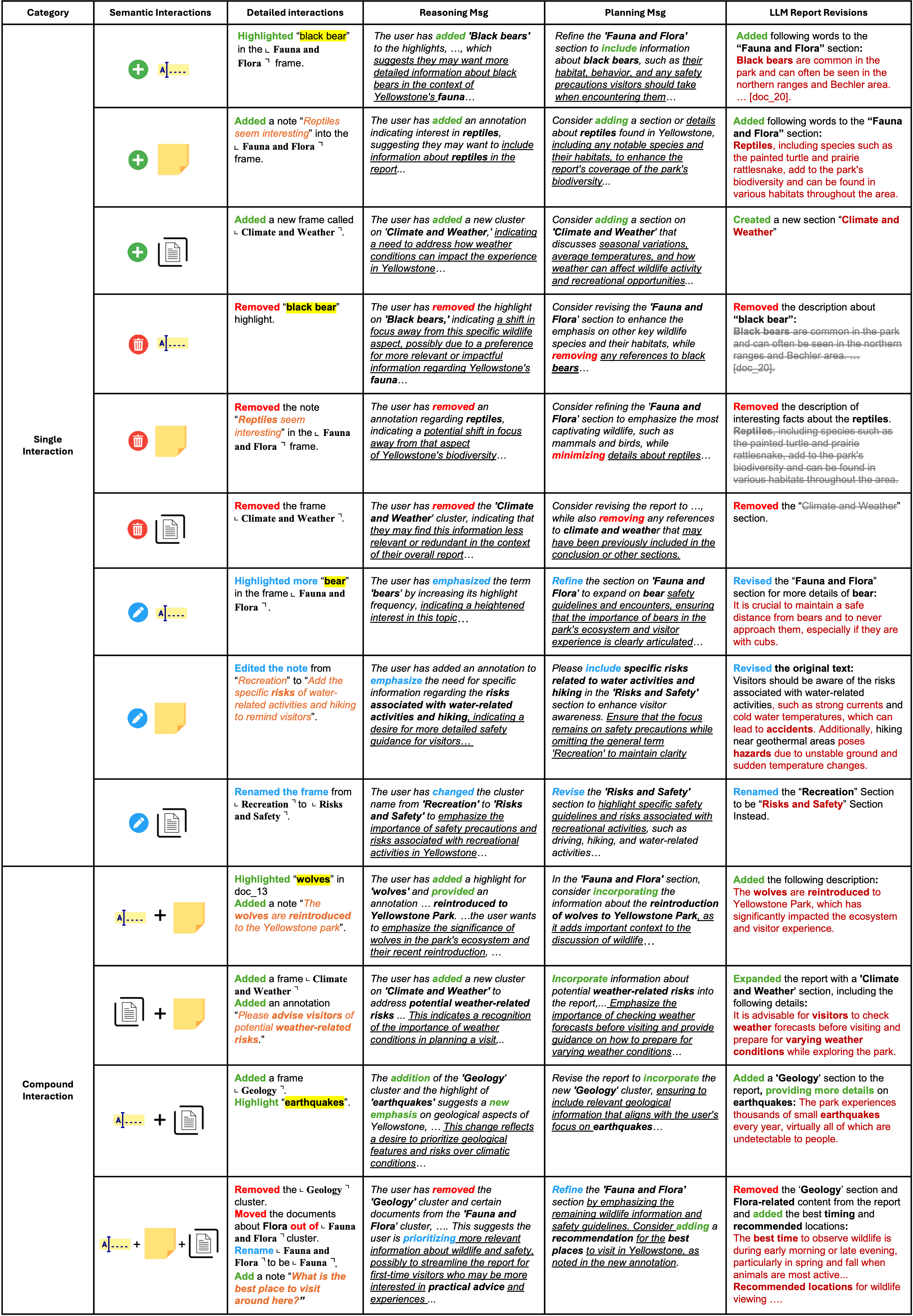}
    \caption{Examples of mapping semantic interactions to final revisions. Corresponding colors and bold text highlight these relationships.
    }
    \label{fig:cs2}
\end{figure}

\subsection{S\_PRISM Refinement Prompts}
\begin{longtable}{|p{0.15\textwidth}|p{0.75\textwidth}|}
\hline
\textbf{Internal Prompts} & \textbf{Prompts} \\
\hline
Interaction Reasoning Prompt \textbf{(\textcolor{blue}{Interaction Inferring Agent}}) & You are an expert in sensemaking. Your task is to reason why user changes the workspace after reading all the documents and give me a concise answer of how the facts in the previous report will change.

I will provide you with:
\begin{enumerate}
    \item The previous workspace report
    \item A detailed diff showing what has changed between the previous workspace and the current one
    \item The current document structure and human visual marks in the workspace
\end{enumerate}

These interactions convey different meanings:
\begin{enumerate}
    \item Highlight: Keywords marked by the user to emphasize importance and guide your attention.
    \item Annotation: The user's personal insights related to a specific document or cluster of information. Do not use it to change the cluster name.
    \item Negative\_highlight: Keywords marked as irrelevant by the user. Omit any terms added to negative\_highlights from new reports. If a term is removed from highlights, you may include it in new reports only if you determine it holds significant value.
\end{enumerate}

Your task:
\begin{itemize}
    \item Reason why user changes the workspace and write prompts for guiding the report refinement. Provide specific, concrete details rather than abstract descriptions.
\end{itemize}
\\ 
\hline
Assistant Prompt for Interaction Reasoning \textbf{(\textcolor{blue}{Interaction Inferring Agent}}) & The returned JSON objects should follow this format:
\begin{lstlisting}[style=JSON]
   {
        "Reasoning": "Reason why user changes the workspace. DO NOT give abstract description.",
        "Command": "You should write prompts for guiding the report refinement in one paragraph. DO NOT give abstract command. Leave the command empty if you think the report does not need to be refined. If the cluster name is changed, you should change the cluster name in the report."
    } 
\end{lstlisting}
\\
\hline
Task Reasoning Prompt \textbf{(\textcolor{fontgreen}{System Inferring Agent}}) & You are an expert in sensemaking. Your task is to reason identify if and understand why user changes the task description and report structure after reading all the documents and give me a concise answer of how the facts in the previous report will change.
I will provide you with:
\begin{enumerate}
    \item The task description and report structure changes
\end{enumerate}
Your task:
\begin{itemize}
    \item Identify the changes in the task description and report structure.
    \item Write prompts for guiding the report refinement based on the changes.
\end{itemize}
\\ 
 
\hline
Assistant Prompt for Task Reasoning \textbf{(\textcolor{fontgreen}{System Inferring Agent}}) & The returned JSON objects should follow this format:
\begin{lstlisting}[style=JSON]
{
    "Reasoning": "Identify the changes in the task description and report structure.",
    "Command": "Write a prompt for guiding the report refinement based on the changes in one sentence."
}
\end{lstlisting} \\
\hline
Refining Prompt \textbf{(Refining Agent)} & You are analyzing visual workspaces that contain clusters of information. I will provide you with:
    \begin{enumerate}
        \item The previous workspace report
        \item A detailed diff showing what has changed between the previous workspace and the current one
        \item A report of human intention of the changes.
    \end{enumerate}

    Your task:
    \begin{itemize}
        \item REVISE content from the previous report only to improve accuracy and conform to the new output format; do not rephrase.
        \item ADD new sections ONLY for newly introduced clusters or elements and revise the introduction and conclusion of the report.
        \item Format any new content to match the style of the existing report
        \item TREAT this as a report that summarizes the workspace. DO NOT include the words such as "new clusters/new highlights reveal that..."
        \item DO NOT include words like "cluster" or "document" in the report.
        \item DO NOT omit any cluster summaries. Even if one cluster includes all documents, summarize it as a standalone group.
    \end{itemize}
    \\
\hline
Assistant Prompt for Refinement \textbf{(Refining Agent)} &The returned JSON objects must follow this format:
\begin{lstlisting}[style=JSON] 
{
    "Introduction": "<Introduction>",
    "Middle_paragraph": [
        {
            "id": "use the original cluster id",
            "name": "If the cluster's original name is empty, must summarize the content as the new name. If not, DO NOT change and use the original name",
            "summarization": "summarize this cluster consist of documents",
            "text": "<ClusterDescription>",
            "name entities": "list important entities in 'text' with document id as citation, like 'name[document_id1, document_id2]'.",
            "keywords": "list important keywords you found in the input JSON cluster with document id as citation, like 'keyword[document1 id, document2 id]'",
            "subclusters": {
                "subcluster_id": {
                    "name": "If the cluster's original name is empty, must summarize the content as the new name. If not, DO NOT change and use the original name",
                    "text": "<SubClusterDescription>",
                },
            }
        },
        {
            ...
        }
    ],
    "Conclusion": "<Conclusion>"
}
\end{lstlisting}
Note you should use cluster and node id provided to you in the input JSON object. Remember there are clusters and possibly sub clusters. DO NOT create your own sub clusters.
\\
\hline
Prompt template for interaction changes  \textbf{(Refining Agent)} &  
1. "Cluster changes: " + <document structure changes>

Template of <document structure changes> (The final JSON data will be cleaned by removing items with empty strings.):
\begin{lstlisting}[style=JSON] 
{
    "removed_clusters": [the id list of deleted clusters],
    "added_clusters": [
        {
            "cluster_id": "",
            "name": "",
        },
        ...
    ],
    "cluster_changes": [
        "name_changes": {
            "old": "",
            "new": ""
        },
        "member_changes": {
            "removed": "",
            "added": "",
        },
        subcluster_changes: {
            "removed": "",
            "added": "",
            "modified": {
                "subcluster_id": string;
                "name_changes": {
                    "old": "",
                    "new": "",
                },
                "member_changes": {
                    "removed": "",
                    "added": "",
                }
            }
        },
        "structural_changes": {
            "old_structure": main' or 'sub',
            "new_structure": main' or 'sub',
            parent_cluster: "",
        }
    ]
}
\end{lstlisting}

2. "Human interaction changes:" + <semantic interactions>

Template of <semantic interactions> (The final JSON data will be cleaned by removing items with empty strings.):
\begin{lstlisting}[style=JSON]
{
    "highlights": {
        "added_items": [],
        "removed_items": [],
        "modified_items": []
    },
    "negative_highlights: {
        "added_items": [],
        "removed_items": [],
        "modified_items": []
    },
    "annotations": {
        "added": {
            "cluster1 id": "added annotation to cluster1",
            ...
        },
        "removed": {
            "cluster2 id": "removed annotation to cluster2",
            ...
        }
    },
}
\end{lstlisting}
 \\
\hline
Prompt template for task changes \textbf{(Refining Agent)} & 
1. "Task description changes: " + <task description changes>

Template of <task description changes>:

\textit{New task description: "the new task description"}

2. "Report structure changes: " + <report structure changes>

Template of <report structure changes>  
\begin{lstlisting}[style=JSON]
{
	"added": [
        {
			"id": "added structure item id",
			"label": "added structure item label",
			"value": "added prompt"
		},
        ...
    ],
	"removed": [
		{
			"id": "removed structure item id",
			"label": "removed structure item label",
			"value": "removed prompt"
		},
        ...
	],
	"modified": [
        {
			"id": "modified structure item id",
			"label": "modified structure item label",
			"value": "modified prompt"
		},
        ...
    ],
	"reordered": false/true
}
\end{lstlisting}

\\
\hline
\end{longtable}

\subsection{Empirical Evaluation Dataset}
\begin{center}
\noindent\begin{minipage}{0.7\textwidth}
\renewcommand{\arraystretch}{1.5}
\begin{tabular}{|>{\centering\arraybackslash}m{2cm}|>{\centering\arraybackslash}m{3.5cm}|>{\centering\arraybackslash}m{6cm}|}
\hline
\rowcolor{gray} \textcolor{black}{Category} & \textcolor{black}{Interactions} & \textcolor{black}{Number of Pairs} \\
\hline
No change & - & 3 \\
\hline
\multirow{3}{=}{\centering Highlight} & Add highlights & 4 (Add 1, 3, 5, and 7 highlights) \\
\cline{2-3}
 & Remove highlights & 4 (Remove 1, 3, 5, and 7 highlights) \\
\cline{2-3}
 & Modify highlight weight & 4 (Modify 1, 3, 5, and 7 highlights) \\
\hline
\multirow{2}{=}{\centering Note} & Add notes & 3 (Add 1, 2, and 3 notes) \\
\cline{2-3}
 & Remove notes & 3 (Remove 1, 2, and 3 notes) \\
\hline
\multirow{4}{=}{\parbox{2cm}{\centering Cluster \\ Reorganization}} & Add clusters & 2 (Add 1 and 2 clusters) \\
\cline{2-3}
 & Delete clusters & 2 (Delete 1 and 2 clusters) \\
\cline{2-3}
 & Adjust cluster members & 3 (Remove 1, 2, 3 documents from one cluster) \\
\cline{2-3}
 & Adjust cluster names  & 3 (Rename 1, 2, 3 clusters) \\
\hline
\multirow{1}{=}{\centering Compound Interactions} & \multirow{1}{=}{\parbox{3.5cm}{\centering Adjust highlights, notes, and clusters}} & 4 (Highlights + notes, Highlights + cluster member changes, Notes + cluster number changes, Highlights + notes + cluster number changes) \\
\hline
\end{tabular}
\captionof{table}{35 pairs of original and interacted workspaces in case study, covering no interaction, few interactions, and compound interactions.}
\label{tab:interactions}
\end{minipage}
\end{center}

\newpage
\subsection{Default Prompts for User Study}
\subsubsection{Phase I}
\begin{longtable}{|p{0.15\textwidth}|p{0.75\textwidth}|}
\hline
\textbf{Prompt Settings} & \textbf{Default Prompts} \\
\hline
Task Description & Imagine you are an FBI agent analyzing the related events. The main task is to predict the nature of the terrorists' threat, including when and where this threat will be carried out, who is involved, and what will happen.\\
\hline
Introduction & Summarize your findings from those documents and clusters as the first paragraph of the Bottom Line Up Front manner. It should include the necessary information of Who, When, Where, and What.\\
\hline 
Cluster Description & [SUMMARIZATION of each cluster first in a sentence] [ANALYSIS CONTENT must include specific information of Who, When, Where and What. Be as detailed as you can, at least three sentences.]\\
\hline
Subcluster Description &Provide analysis of sub-clusters within each main cluster, including specific relationships, patterns, and insights that emerge from the sub-cluster analysis.\\
\hline
Conclusion & Draw conclusions of how those clusters connect to each other, inferences, and speculate on causes and effects. Do not re-list the events mentioned here.\\
\hline
\end{longtable}
\subsubsection{Phase II}
We left the task description prompt blank to allow users to specify their own requirements.
\begin{longtable}{|p{0.15\textwidth}|p{0.75\textwidth}|}
\hline
\textbf{Prompt Settings} & \textbf{Prompts} \\
\hline
Introduction & Provide a "Bottom Line Up Front" (BLUF) executive summary in one paragraph. Answer: "What should a first-time visitor to Yellowstone know?" Synthesize the most critical findings about planning, safety, attractions, and logistics into actionable takeaways.\\
\hline 
Cluster Description & For each major topic (attractions, logistics, safety, etc.), provide: 1) One-sentence overview, 2) Detailed analysis covering What (key features/activities), Where (specific locations), When (best timing/seasons), and Who (target visitor types).\\
\hline
Subcluster Description & Analyze subtopics within each main category. Highlight specific relationships, practical considerations, and insider insights that enhance the visitor experience.\\
\hline
Conclusion & Synthesize how different aspects of Yellowstone connect to create an optimal visit strategy. Focus on: 1) How timing, location choices, and activities interconnect, 2) Key decision points for first-time visitors, 3) Potential challenges and how to mitigate them.\\
\hline
\end{longtable}

\end{document}